\documentclass[12pt]{article}
\usepackage{epsfig}
\usepackage{amssymb}

\newcommand{\be}{\begin{equation}}
\newcommand{\ee}{\end{equation}}
\newcommand{\bea}{\begin{eqnarray}}
\newcommand{\eea}{\end{eqnarray}}
\newcommand{\ov}{\overline}
\newcommand{\ve}{\varepsilon}

\newcommand{\ba}{\begin{array}}
\newcommand{\ea}{\end{array}}

\def\la{\mathrel{\mathpalette\fun <}}

\def\fun#1#2{\lower3.6pt\vbox{\baselineskip0pt\lineskip.9pt
\ialign{$\mathsurround=0pt#1\hfil##\hfil$\crcr#2\crcr\sim\crcr}}}

\title{Production of light antinuclei in heavy  ion collisions}
\author{{B.L. Ioffe, I.A. Shushpanov and K.N. Zyablyuk}\\
{\it \normalsize Institute of Theoretical and Experimental Physics},\\
{\it \normalsize B.Cheremushkinskaya 25, 117218 Moscow,Russia}\\
{\tt \normalsize ioffe@itep.ru, shushpan@itep.ru, zyablyuk@itep.ru}}
\date{}

\begin{document}
\maketitle

\vspace{10mm}

\begin{abstract}

The antideuteron and antihelium-3 production rates at high-energy
heavy ion collisions are calculated in the framework of fusion
mechanism. It is supposed, that
$\bar{p},\bar{n},\bar{d},~^3\ov{He}$ participating in the
fusion are moving in the mean field of other fireball
constituents. It is demonstrated that at high energies, where many
pions are present in the fireball, the number of produced
$\bar{d}$ and $^3\ov{He}$ is determined by the balance between
created and desintegrated (mainly in collisions with pions)
$\bar{d}$ and $^3\ov{He}$. The explicit formulae for coalescence
parameters are presented and compared with the data.

\end{abstract}

\bigskip

\centerline{PACS numbers:  25.75.-q, 25.75.Dw, 12.38.Mh}

\vspace{10mm}

\section{Introduction}

In recent experiments on heavy ion collisions the production of
 light antinuclei --- antideuterons and antihelium-3 --- was measured
 \cite{1},\cite{2},\cite{3},\cite{4}. Once may expect, that this
 process proceeds at the intermediate stage of the evolution of
 the fireball, created in heavy ion collisions. Because of their
 small binding energy antinuclei are formed at the stage, when the
 hadrons are already formed, the density of hadronic matter is of
 order of normal nuclear density, and the particle collisions are
 still important. We will call this stage as the "dense gas"
 stage of fireball evolution. It is important to have a theoretical
 description of antinuclei production in heavy ions collision,
 because the dense gas stage is evidently after the stage, where
 quarks and gluons transform into hadrons, but on the other
 hand this stage precedes the last stage of the fireball evolution
 (sometimes called the thermal freeze-out), when hadronic spectra
 are formed. The theoretical calculations of antinuclei production
 and their comparison with experiment can shed some light on the
 dense gas stage of the fireball evolution and allows one to do
 one step back from the final stage, about which we have direct
 information from experiment. Particularely, as it will be shown,
 the volume of the fireball and the hadron densities  at the dense
 gas stage can be estimated  from the data on antinuclei production.
 It is a common belief, that the antinuclei production proceeds as a
 fusion process:
 \be
 \bar{p} \, + \, \bar{n}\,  \to \, \bar{d}
\label{1}
\ee
in case of antideuterons and
\be
\bar{p}\,+\,\bar{p}\,+\,\bar{n} \, \to \, ^3 \ov{He}
\label{2}
\ee
in case of $^3\ov{He}$. According to the dominant coalescence
mechanism (\ref{1},\ref{2}) it  is convenient to characterize
$\bar{d}$ and $^3\ov{He}$ production by the coalescence parameters
\be
B_2\, =\,E_{\bar{d}} \,{d^3N_{\bar{d}} \over d^3p_{\bar{d}}}
 \left(  E_{\bar{p}} {d^3N_{\bar{p}} \over d^3p_{\bar{p}}} \right)^{-2},
 \qquad p_{\bar p}\,=\,p_{\bar d}/2 \; ,
\label{3}
\ee
\be
B_3\, =\,E_{\ov{He}}\, {d^3 N_{\ov{He}} \over d^3p_{\overline{He}}}
\left( E_{\bar{p}} {d^3N_{\bar{p}} \over d^3p_{\bar{p}}} \right)^{-3},
\qquad p_{\bar p}\,=\,p_{\ov{He}}/3 \; ,
\label{4}
\ee
where one assumes $d^3N_{\bar{n}}/d^3p_{\bar n}=d^3N_{\bar{p}}/d^3p_{\bar{p}}$. In what
follows we consider only the central heavy ions collisions.

The basic ideas  of our approach are the following. We assume,
that the coalescence mechanism (\ref{1},\ref{2}) is the dominant one in the
production of heavy nuclei. (For $\bar{d}$ production it will
be shown by direct calculation, that the contribution of competing
process $\bar{p}+\bar{n}\to \bar{d}+\pi$ is small, see Appendix B.)
The fusion reactions (\ref{1},\ref{2}) cannot proceed if all particles are on
 mass shell. However, in the fireball at the dense gas of its evolution,
$\bar{p},\bar{n},\bar{d},~^3\ov{He}$ are not on mass shell,
since they interact with surrounding matter. One may consider
their movement as a propagation in the mean external complex field
caused by the matter. The interaction with this field leads to
appearance of the mass shifts and widths of all particles
propagating in the medium (or width broadening for unstable ones),
analogous to refraction and attenuation indeces in the case of
photon propagation. Another important ingredient of our approach
is the balance of antinucleous production and desintegration
rates. This balance is achieved because of large density of pions
in the fireball and high rate of $\pi +$ ($\bar{d}$ or $^3\ov{He}$)
collisions leading to antinucleous
desintegration.  The balance does not imply a statistical
equilibrium, but rather a stationary process. The statistical or
thermal equilibrium are not assumed in the calculation.

In most of previous investigations the production of deuterons or light nuclei was considered,
but not the production of antinuclei. These processes have some common features,
but also some differences.

The production of light nuclei in heavy ion collisions nas been studied
for many years (for an early review see \cite{CK}). The first calculation of the
deuteron production rate was performed in \cite{BP} within simple model
of optical potential in nuclei, resposible for the deuteron formation.
In \cite{SZ} it was proposed that the proton and neutron bind together
if their relative momentum is less than some $p_0$, a phenomenological parameter
to be determinad from experiments. Then $B_2= {2 \over E_p} \cdot {3\over 4} \cdot {4\pi\over 3}p_0^3$
is proportional to the propability to find the proton and neutron inside the sphere of radius $p_0$ in momentum
space. A thermodynamical approach to the deuteron formation was applied in
\cite{M}, where the deuterons were assumed to be in thermodynamical equlibrium
with other nucleons in the fireball.

In \cite{BJKG} the sudden approximation of quantum mechanics was
first applied to the light nuclei formation. It implies short transition from the high-density
stage $|i\rangle $ consisting of the protons and neutrons only, to the low-density stage $| f \rangle $,
consisting of the deuterons as well. The amplitude of the deuteron formation is given by the overlap
of the wave finctions $\langle f| i \rangle$. Under simplifying assumptions that these states are
free moving particles, uniformly distributed in a box of  volume $V$ and neglecting the deuteron size, one
obtains the coalescense parameter:
$$
B_2\,=\, {2\over E_p} \,\cdot \, {3\over 4} \, \cdot\, {(2\pi)^3\over V}
$$
where $3/4$ is the ratio of the spin weights, the factor $2/E_p$ comes from invariant definition
of the phase space element $d^3p/E$.

Most of investigations of the light cluster formation in heavy ion collisions
represent extentions of the sudden approximation result
for variously prepared initial state (sourse). Among them are different sourse geometries \cite{SY}, \cite{Mr},
expanding systems \cite{DHSZ}, intranuclear cascade model \cite{GFR}, diagrammatic approach
\cite{DPV}, relativistic quantum molecular dynamics model \cite{NKKSM}, Boltzmann-Uehling-Uhlenbeck
transport model  \cite{CKL}. In most cases the results are based on computer simulations and
analytical expessions are unavailable.

A novel phenomenon is observed if one tries to go beyond the sudden aproximation.
In \cite{Kap} the deuteron formation was considered in the time-dependent perturbation theory,
If the interaction, responsible for the deuteron formation, is switched on during some finite time
$\tau$, the coalescense parameter gets multiplied on additional factor
$\sim (\varepsilon\tau)^{1/2}$, where $\ve$ is the deuteron binding energy. Exact value
of $\tau$ is not determined; as argued in \cite{Kap}, it should be of order of colliding
nuclei radius, so $(\ve\tau)^{1/2}\sim 1$.

In all mentioned above approaches the interaction of the deuterons with the fireball
enviroment was not accounted. As it will be shown below, this interaction, especially with pions,
plays an essential role. As a consequence the coalescence parameter
acquires the factor $(\ve/\Gamma)^{1/2}$, where $\Gamma$ is the
deuteron in-medium width, which is calculated here. According to our calculation
$(\ve/\Gamma)^{1/2}\sim 1/10$.

Another transport model of the deuteron formation in nuclear collisions is available in
literature \cite{DB, BSKR}. It was assumed, that the deuteron formation and desintegration
goes mainly via triple nucleon reaction $NNN\leftrightarrows dN$. At low (nonrelativistic)
energies the reaction rates were calculated using impulse (Born) approximation
with the help of Faddeev equation. Again, the thermodynamical equilibrium was assumed
in the calculations. This approach also does not account the deuteron desintegration by the
pions.

It must be stressed that we apply our approach only to the process of {\it anti}deuteron
(or antihelium) formation in high-energy central heavy ion collisions. In this case, prior
to antinuclei formation, the hadrons experience at least few collisions with each other
and the concept of the hot dense gas seems reasonable to apply.
Although the approach is valid for the deuteron (or helium) formation processes,
the results should be used with care: a part of the outgoing deuterons may
be just some fragments of primary colliding nuclei.
Is is rather difficult to separate this part from the deuterons, synthesized from the hot
hadronic gas, at least some experimental cuts should be applied.
We will not consider this point here, restricting ourselves only by antinuclei processes.

The material of the paper is presented as follows. In Section 2 the effective low energy
Lagrangian for fusion processes (\ref{1},\ref{2}) is constracted. The value of
effective coupling constant is found by two ways: 1) by
considering the elastic scattering amplitude $\bar{p}+\bar{n}\to
\bar{p}+\bar{n}$ (in case of $\bar{d}$ production); 2) by
consideration of $\bar{d}$ or $^3\ov{He}$ polarization
operators, on the base of the field theory. With the help of effective
Lagrangian the cross sections of fusion reactions are
calculated. In Section 3 the transport equation for antinucleous  production and
propagation  in the fireball is formulated. The formulae for the
widths and mass shifts of particles, moving in the medium are
presented and the formation rates of $\bar{d}$ and
$^3\overline{He}$ are calculated in terms of  effective
$\bar{p},\bar{n},\bar{d},~^3\overline{He}$ widths in the medium.
In Section 4 the balance conditions for $\bar{d}$ and
$^3\overline{He}$ production are formulated and the explicit
formulae for coalescence parameters (\ref{3}),(\ref{4}) are
presented. Section 5 is devoted to comparison with experimental
data. The model of pion and nucleon densities at the dense gas
stage of the fireball evolution is formulated. With the help of this
model the values of the widths were calculated and it is
demonstrated, that all assumptions, used in the calculation are
fulfilled. The coalescence parameters are calculated numerically
for experimental conditions and compared with the data. Section 6
presents our conclusion. The  details of the calculation are given
in the Appendices.

\section{Effective low energy Lagrangians and fusion cross sections.}

\begin{figure}[tb]
\hspace{60mm} \epsfig{file=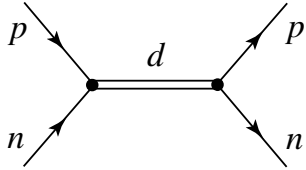, width=40mm}
\caption{Elastic $pn$ scattering through $d$ formation}
\label{fig_pnd_s}
\end{figure}

At first let us consider the antideuteron production. Evidently,
$\sigma_{\bar{p}+\bar{n}\to\bar{d}} =\sigma_{p+n\to d}$. The low
energy effective interaction Lagrangian has the form
\be
L\,=\,g\psi_p\psi_n \varphi^+_d \, ,
\label{5}
\ee 
where $\psi_p$, $\psi_n$ and $\varphi_d$ are nonrelativistic $p,n$ and $d$  wave functions and
spin effects are neglected. Let us temporarely  suppose, that $pnd$
interaction is point-like, i.e. neglect the nuclear force radius
$r_0$. Following Landau \cite{10}, calculate the forward $p+n\to p+n$ 
elastic scattering amplitude. In this approximation it
corresponds to Fig.\ref{fig_pnd_s} diagram. According to standard rules  
the relativistic amplitude is given by:
\be
\label{6}
M\,=\,{-g^2\over s-m_d^2} \,\approx \, {-g^2\over 4m(E+\ve)}
\ee
where $s=(p_p+p_n)^2$, $m$ is nucleon mass, $\ve=m_p+m_n-m_d=2.225\,{\rm MeV}$
is the deuteron binding energy, $E$ is the c.m.~energy, $\sqrt{s}=m_p+m_n+E$. 

On the other hand, the nonrelativistic $s$-wave amplitude $f$ of elastic scattering 
 according to Bethe-Peierls \cite{11} theory is equal
\be
f_0\,=\,{1\over k({\rm ctg}\,\delta_0 - i)}\,\approx\,{i\over k-i\alpha} \, ,
\label{7}
\ee 
where $\alpha=\sqrt{m\ve}$,  $k$ is the proton (or neutron) momentum in c.m.~frame. 
At  the deuteron pole, $s\to m^2_d$, $k\to i\alpha$, the amplitudes (\ref{6}) and
(\ref{7}) must be equal (keeping in mind the different normalizations
 $M=8\pi\sqrt{s} f$). This requirement determines the
value of the effective coupling constant \cite{10}:
\be
 g^2\,=\,128\pi\, m\sqrt{m\ve}
\label{8}
\ee 
It must be stressed, that $g^2\to  0$ at $\ve\to 0$, which means that the fusion
rate goes to zero at $\ve\to 0$. Physically this is natural: the probability of 
production of large size antideuteron in heavy ion collision decreases with increasing of
its size. It is an easy task to account for the finite value of nuclear forse radius $r_0$, 
see \cite{12} and Appendix A. As a result,  the forward
scattering amplitude $f_{0}$ at the pole $k=i\alpha$ gets an
additional factor $(1-\alpha r_0)^{-1}\approx 1.67$. With account of
this correction $g^2$ becomes equal:
\be
g^2\, =\, 128 \pi\, m\sqrt{m\ve}\,(1-\alpha r_0)^{-1}
\label{9}
\ee
The invariant cross section for the fusion process $pn\to d$ is found by common rules:
\be
E_d{d^3 \sigma_{pn\to d}\over d^3 p_d}\, =\, {3\over 4} \cdot 
 {\pi g^2\over 4\sqrt{(p_n p_p)^2-m^4}}  \, \delta^4(p_p+p_n-p_d),
\label{10}
\ee 
where $E_p,E_n$ and $E_d$ are $p$, $n$ and $d$ total energies, 3/4 is the 
ratio of the spin weights. In fact, the cross-section (\ref{10}) is valid only for 
low (nonrelativistic) cm energies. At high energies excited states are produced
(for instance $N\Delta$ resonance) which may decay into the antideuteron. 
Here wo do not consider these rather complicated (in medium) processes,
assuming they are small enough.

Let us now derive the expression for the effective coupling
constant $g^2$ (in the limit $r_0\to 0$) by another method -- by
 consideration of the deuteron Green function and polarization
operator in field theory. This approach, of course, gives the same
result as above in case of deuteron, but it will be useful in
consideration of $^3\ov{He}$ production, where the scattering 
theory is rather complicated.

For nonrenormalized deuteron Green function $D(p^2)$  we have the
Schwinger-Dyson equation:
\be
\left[\, p^2\,-\,m^2_{d,0}\, -\,\Pi(p^2)\,\right] D(p^2)\, =\,1 \, ,
\label{11}
\ee 
where $m^2_{d,0}$ is deuteron bare mass. After mass renormalization we get
$$
\left\{\,p^2\,-\,m^2_{d,0}\, -\,\Pi(m^2_d)\,- \left[\,\Pi(p^2)\,-\,\Pi(m^2_d)\, \right]\right\} D(p^2)\,=\,1\, ,
$$
\be
m^2_{d,0}\, +\,\Pi(m^2_d) \, =\,m^2_d\, ,
\label{12} 
\ee
where $m_d$ is the physical deuteron mass. Then we perform the Green
function renormalization $D(p^2)=Z_2 D_{ren}(p^2)$ so that
\be
D_{ren}(p^2)_{p^2\to m^2_d}\,\to\, {1\over p^2-m^2_d}
\label{13}
\ee
\begin{figure}[tb]
\hspace{30mm} \epsfig{file=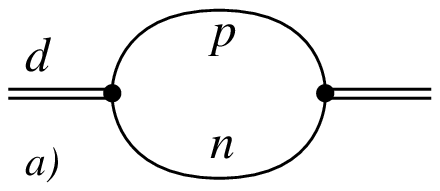, width=40mm}
\hspace{20mm} \epsfig{file=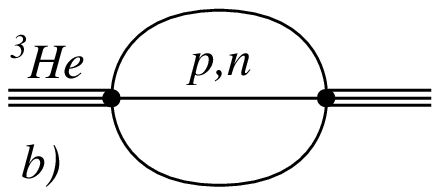, width=40mm}
\caption{Deuteron and helium-3 self energies}
\label{fig_pn_l}
\end{figure}
The main assumption in this approach, used in derivation of the Bethe-Peierls equation
(\ref{7}), is: vertex corrections and corrections to nucleon
propagators are neglected, $d$ is assumed to be the bound state of
$p$ and $n$. The polarization operator is contributed only by the
diagram of Fig.~\ref{fig_pn_l}a) and we can write $\Pi=g^2_0\tilde{\Pi}$, where
$g^2_0$ is the nonrenormalized coupling constant, related to 
renormalized one $g^2$ as:
\be
g^2_0\,=\, Z^{-1}_2 g^2 \, .
\label{14}
\ee 
Representing $\Pi$ in (\ref{12}) by  dispersion relation we get
\be
Z_2\left\{\, (p^2-m^2_d)\,+\, {g^2\over Z_2}(p^2-m^2_d) {1\over\pi}
\int ds {Im \,\tilde{\Pi}(s)\over (s-m^2_d)(s-p^2)}\,\right\}
D_{ren}(p^2)\,=\,1
\label{15}
\ee 
The condition of (\ref{13}) results in:
\be
Z_2\, +\,{g^2\over \pi}\,\int ds {Im \,\tilde{\Pi}(s)\over (s-m^2_d)^2}\,=\,1
\label{16}
\ee
If deuteron is the bound state of proton and neutron, then the contribution of
bare deuteron to Schwinger-Dyson equation vanishes and $Z_2=0$. Putting
$Z_2=0$ in (\ref{16}) we get the equation for $g^2$:
\be
{g^2\over \pi}\, \int ds \,{Im \,\tilde{\Pi}(s)\over (s-m_d)^2}\,=\,1
\label{17}
\ee
$Im \,\tilde{\Pi}(s)$ can be easely calculated for Fig.~\ref{fig_pn_l}a) diagram
(nonrelativistic approximation can be used). As a result the same
expression for $g^2$ (\ref{8}) as in the first method is obtained.

Let us turn now to the study of $^3\ov{He}$ production. The effective
nonrelativistic Lagrengian has the form
\be
L\,=\,G_0\psi_{p_1}\psi_{p_2}\psi_n\psi^+_{He}
\label{18}
\ee 
The derivation, performed above in the second approach, can be repeated
and eq.~(\ref{17}) with substitution $g^2\to G^2$, $m_d\to m_{He}$
is found. The $Im \,\tilde{\Pi}(s)$ is described now by the diagram
of Fig.~\ref{fig_pn_l}b). There is, however, the essential difference in
calculation of the Fig.~\ref{fig_pn_l}b) diagram contribution to (\ref{17}) in
comparison with that of Fig.~\ref{fig_pn_l}a): the integral in (\ref{17}) is
linearly ultraviolet divergent in nonrelativistic approximation.
(This divergence still persists in relativistic calculations.)
This circumstance physically corresponds to the well known fact,
that in nuclear physics the 3-body problem with $\delta$-function
potential cannot be correctly formulated: it is necessary to have
an additional information about the  interaction at small
distances. For our purposes  it is sufficient to have a rather crude
estimation of $G^2$. So, we put ultraviolet cut-off in
the integral in (\ref{17}). The calculation gives:
\be
G^2\, =\, 36 \sqrt{3}\,(4\pi)^3 \, {m\over \Lambda}
\label{19}
\ee 
In numerical calculations $\Lambda=300\,{\rm MeV}$ will be taken.

\section{Transport equation. Calculation of collisions integral.}

As mentioned in the  Introduction, we suppose, that antideuterons
and antihelium-3 are formed at the dense gas stage of the fireball
evolution, which follows after "chemical freeze-out" stage
\cite{13,14}. Let us assume, that particle propagations at this stage can
be described by means of classic transport (kinetic) equations.

For definiteness, consider first the $\bar{d}$-production. We will use the
notation $q_i(x,p)$, $i=\bar{p},\bar{n},\bar{d},\pi, ...$ for the
double densities in coordinate and momentum spaces and
$n_i(x)=\int q_i(x,p)\, d^3p$ for the coordinate densities. ($q_i(x,p)$ are Lorentz
invariant.) Let us choose the c.m.~frame of colliding ions. The
transport equation for $q_{\bar{d}}(x,p)$ has the form:
$$
{m_d\over E_{\bar{d}}} \, {\partial q_{\bar{d}}(p_{\bar{d}},x)\over\partial x_{\mu}}u^{\bar{d}}_{\mu}\,
 =\, {\partial q_{\bar{d}}\over \partial t}\, +\, {\bf v}_{\bar{d}}
\bigtriangledown q_{\bar{d}}\, = \hspace{49mm}
$$
\be
=\, \int d^3 p_{\bar{p}}\, d^3 p_{\bar{n}}\, q_{\bar{n}}(p_{\bar{p}})
 \, q_{\bar{n}}(p_{\bar{n}}) \, \omega_{\bar{p}\bar{n}\to \bar{d}} \,
-\,q_{\bar d}(p_{\bar d}) \sum_i \int d^3 p_i \, q_i(p_i) \, \omega_{{\bar d}i\to X}
\label{20}
\ee
where $u^{\bar{d}}_{\mu}=(1,{\bf v}_{\bar{d}})/\sqrt{1-v^2_{\bar{d}}}$
is antideuteron 4-velocity, $\omega_{\bar{p}\bar{n}\to \bar{d}}$ is the fusion reaction
 propability, proportional  to differential cross section
\be
\omega_{{\bar p}{\bar n}\to {\bar d}} \, =\, {\sqrt{(p_{\bar p} p_{\bar n})^2-m^4}\over
E_{\bar p} E_{\bar n}} \, {d^3\sigma_{{\bar p}{\bar n}\to {\bar d}} \over d^3 p_{\bar d}}
\label{omsig}
\ee
and similarly for the desintegration process ${\bar d}i\to X$
due to collisions of $\bar{d}$ with $i$-th constituent of the fireball
($i=\pi,K,p,n$, etc.).  The terms, where $\bar{d}$ appears in the momentum interval
$p_{\bar{d}}$, $p_{\bar{d}}+\Delta p_{\bar{d}}$ as  a result of
elastic collisions are neglected. The  cross section
$\sigma_{\bar{p}\bar{n}\to \bar{d}}$ is defined by (\ref{10}), the
cross sections of $\bar{d}$ collisions  with fireball constituents
$(\sigma_{\pi\bar{d}}$, ...) can be found from the experimental
data. Necessary applicability conditions of (\ref{20}) is that
particles wave lengths $\lambda_i=p^{-1}_i$ should be much less,
than the mean distances $d$ between fireball constituents,
$\lambda_i\ll d$.

At the dense gas stage of the fireball evolution all particles
inside the fireball should be considered as moving in the mean
field of other fireball constituents. As a consequence, the masses
 are shifted in comparison with their vacuum values. Similarly,
 due to interaction with medium constituents, the widths
 $\Gamma_i$ appear (or width broadening, if the particle has its
 proper widths). The mass shift $\Delta m(E)$ and width
 $\Gamma(E)$ are expressed in terms of the forward scattering
 amplitude $f(E)$ of the particle on $i$-th medium constituents
 (see\cite{15},\cite{16} and references therein)
 \bea
 \Delta m(E)& = & -2\pi\, {n_i\over m}\, Re\,f_i(E)  \nonumber \\
 \Gamma(E)& = & 4\pi\, {n_i\over m}\, Im \,f_i(E)\,=\,{n_i p\over m}\, \sigma_i(E)\, , \label{22}
\eea
where $E,p$ and $m$ are particle energy, momentum and mass, $n_i$
is the density of $i$-th constituent in medium.
Eq's.(\ref{22}) take place in the system, where the
constituents are at rest. In the case of moving constituents the
corresponding Lorentz boost  must be performed. (By definition $\Delta
m$ and $\Gamma$ are Lorentz  invariant, for details see \cite{17}).
The relations (\ref{22}) must be summed over all fireball constituents.

The applicability  conditions  of eq. (\ref{22}) are the
following \cite{15,16}:
\def\theenumi{\roman{enumi})}
\begin{enumerate}
\item  $\lambda \ll d$
\item  $| f | \ll d$
\item  The main part of the scattering proceeds at small angles, $\theta\ll 1$.
\end{enumerate}
The conditions i) and iii) are well satisfied  in the cases under
consideration. The condition ii) is fulfilled  not quite well and
some corrections should be taken into account (see below).

Therefore, $\bar{p},\bar{n}$ and $\bar{d}$ can be considered as
Breit-Wigner resonances with varying  masses distributed according to
Breit-Wigner formula. In the process of the fireball  expansion
these Breit-Wigner resonances smoothly evolve to their stable
counterparts. So, we substitute (\ref{10}) in the first term in
the right-hand side of eq.~(\ref{20}) and integrate over
the masses $m'$ of Breit-Wigner resonances. Using the notation $I/E_{\bar d}$
for this term we get: 
\bea
 {I\over E_{\bar d}} & = & \int \,d m_{\bar p}' \,d m_{\bar n}' \, d m_{\bar d}'  \,
 {\Gamma_{\bar p}/2\pi \over (m_{\bar p}'-m_{\bar p})^2+\Gamma_{\bar p}^2/4}\,
 {\Gamma_{\bar n}/2\pi \over (m_{\bar n}'-m_{\bar n})^2+\Gamma_{\bar n}^2/4}\,
 {{\tilde\Gamma}_{\bar d}/2\pi \over (m_{\bar d}'-m_{\bar d})^2+{\tilde\Gamma}_{\bar d}^2/4} \nonumber \\
 & & \times \,{3\pi\over 16}\, {g^2\over E_{\bar d}'}\, \int
 {d^3p_{\bar p}\over E_{\bar p}'} \, {d^3p_{\bar n}\over E_{\bar n}'} \,
 q_{\bar p}(p_{\bar p})\,q_{\bar n}(p_{\bar n}) \,
 \delta^3(p_{\bar p}+p_{\bar n}-p_{\bar d}) \, \delta(E_{\bar p}'+E_{\bar n}'-E_{\bar d}')
\label{ibw}
\eea
where $E_{p}'=\sqrt{m^{'2}+p^2_p}$ etc. Assume, that
$\Gamma_{\bar{p},\bar{n}}$ is much smaller than $m$,
 and the variation of $q_{\bar{p},\bar{n}}(p)$ on the interval $\Gamma$ is also small
 enough, $\Gamma dq/dp \ll q$. (In fact,
 $\Gamma_{\bar{p}}=\Gamma_{\bar n}\approx 200\,{\rm MeV}$, see below and
 $\Delta m\approx 30\,{\rm MeV}$.) Then the distributions
$q_{\bar{p}}(p_p)=q_{\bar{n}}(p_n)$ can be taken out from the
integral sign at the values $p_{\bar{p}}=p_{\bar{n}}=p_{\bar{d}}/2$.
 With a good accuracy we can put $\Gamma_{\bar p}=\Gamma_{\bar n}\equiv\Gamma$.
But  $\tilde{\Gamma}_{\bar d}$  in (\ref{ibw}) generally is not equal to the antideuteron  in-medium
width  $\Gamma_{\bar d}\approx 2\Gamma$. The reason is that $\bar{p}\bar{n}$
 system with $\bar{d}$ quantum numbers at high excitations does not necessarily
 evolve to $\bar{d}$ in the process of the fireball expansion, but can
 decay in other ways. One may expect $\tilde{\Gamma}_{\bar{d}} <\Gamma_{\bar{d}}$.
We keep the ratio $a=\tilde{\Gamma}_{\bar{d}}/\Gamma_d$
 as a free parameter in the calculation. As can be seen below, the
 results weakly depend in this ratio. The calculation of the integral
 $I$ is straightforward and leads to
\be
 I \, =\, {3 \pi^2\over 8 }\,  g^2 \,
\sqrt{ \Gamma (1+a) \over 2 m} \, q^2_{\bar{p}} (p_{\bar{p}})
\label{24}
 \ee

The contributions of direct processes $\bar{p}+\bar{p}\to
\bar{d}+\pi^-,$ $\bar{n}+\bar{n}\to \bar{d}+\pi^+$ and
$\bar{p}+\bar{n}\to \bar{d}+\pi^0$ are small, they comprise not
more than 20\% alltogether  and may be neglected  within the
accuracy of our calculation, see Appendix B.

We restrict  ourselves to consideration of low and intermediate
transverse momentum $p_{p_{\perp}}\la 1$ GeV. At  higher $p_{\perp}$
many phenomena go into the play: radial and elliptic flow, steep  decreasing of
spectrum with increasing of $p_{\perp}$ etc. The consideration of
these effects requires more  refined treatment of the problem.

Turn now to the calculation of the second term in the r.h.s.~of
the transport equation (\ref{20}), corresponding to $\bar{d}$ desintegration rate.
The probability $\omega_{{\bar d}i\to X}$ is proportional to the total
cross section $\sigma_{{\bar d}i\to X}$ in the same way as (\ref{omsig}). So one
may write the disintegration term in (\ref{20}) as:
\be
 \sum_i \int d^3 p_i \, q_i(p_i) \, \omega_{{\bar d}i\to X}\,=\,{m_d\over E_{\bar d}}\,\Gamma_{\bar d}
\label{25}
\ee
where $\Gamma_{\bar d}$ has an obvious interpretation of the antideuteron width. It is equal to:
\be
\label{gammad}
\Gamma_{\bar d}\,=\,{1\over m_d} \sum_i \int {d^3 p_i\over E_i} \, q_i(p_i) \,
\sqrt{(p_{\bar d} p_i)^2-m_d^2m_i^2} \, \sigma_{\bar{d}i\to X}
\ee
It is Lorentz invariant generalization of the width (\ref{22})
and depends on the momentum $p_{\bar d}$ of the antideuteron,
moving in medium. The process of ellastic $\pi {\bar d}$  scattering was neglected in
transport equation.

It must be mentioned, that eq.~(\ref{25}) corresponds to gaseous
approximation, when the screening corrections are not accounted.
In fact, the pion densities in the fireball at SPS or RHIC
energies are such, that the account of screening corrections is
necessary. Such calculation, based on Glauber theory, is presented
in Appendix C. As a result, it was found that the Glauber
correction at SPS or RHIC conditions reduces $\Gamma_p$
approximately by 30\%.

\section{The balance condition. The formula for coalescence parameter for
 {\boldmath $\bar{d}$}.}

Suppose, that the rate of antideuteron collisions with other
constituents of the fireball resulting in antideuteron
desintegration is much larger, than the rate of the fireball
expansion. This happens in case of heavy nucleous collisions at
high energies, when the fireball size  at the dense gase stage is
large, because of large number of produced pions per nucleon. The large
relative pionic density in the fireball leads to high desintegration
rate of antideuterons in collisions with pions. (The collisions
with nucleons are less essential  in $\bar{d}$ desintegration.) In
this case one may expect a balance: the antideuteron production rate 
  (first term in the r.h.s. of (\ref{20})) is equal
to its desintegration rate (second term in r.h.s. of (\ref{20})).
The balance condition determines $\bar{d}$-density:
\be
q_{\bar{d}}(p_d)\,=\,{I\over \Gamma_{\bar{d}} m_d}\,=\,
{3\pi^2g^2\over 32\, m}\,\sqrt{1+a\over \Gamma m} q^2_{\bar{p}}(p_p)
\label{30}
\ee
 The momentum distribution $d^3 N_{\bar{d}}/d^3 p_{\bar{d}}$ entering in (\ref{3})
is obtained from (\ref{25}) by integration over the fireball volume
\be
{d^3 N_{\bar{d}}(p_{\bar{d}})\over d^3 p_{\bar{d}}}\,=\,\int d^3x\, q_{\bar{d}}(p_{\bar{d}},x)
\label{31} 
\ee

\begin{figure}[tb]
 \hspace{50mm} \epsfig{file=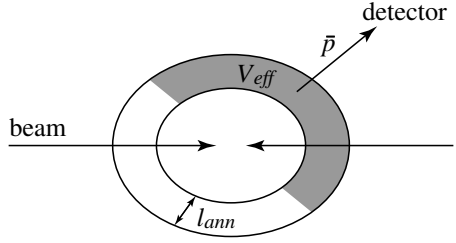, width=60mm}
 \caption{Fireball at the dense gas stage. The effective volume $V_{eff}$ is the half of the
 outer shell of thikness $l_{ann}$, from which the antiprotons reach detector}
 \label{fig_fire_b}
 \end{figure}

From equations (\ref{3}),(\ref{30}),(\ref{31}) and (\ref{9})  we find the
 coalescence parameter
\be
B^{th}_2 \, =\, {24 \pi^3\over E_{\bar{p}}} \times 1.67 \,
\sqrt{\ve (1+a)\over 2\Gamma}  \,{\int d^3 x \, q^2_{\bar{p}}(p_{\bar{p}},x)\over 
[\int d^3 x\, q_{\bar{p}}(p_{\bar{p}},x)]^2}
\label{32} 
\ee
 Since the $x$-dependence of $q_{\bar{p}}(p_{\bar{p}},x)$ is not known,
we replace (\ref{32}) by:
\be
B^{th}_2 \, =\, {24 \pi^3\over E_{\bar{p}}} \times 1.67\,
\sqrt{\ve(1+a)\over 2\Gamma} \, {2\over V}\, {\ov{n^2}_{\bar{p}}\over (\bar{n}_{\bar{p}})^2}
\label{33}
\ee
where $V$ is the fireball volume, $\bar{n}_p$ and $\ov{n^2}_p$ are 
the mean and mean square $\bar{p}$ densitites in the fireball. (The coordinate dependence of
$\sqrt{\Gamma}$ is neglected). $B^{th}_2$  is Lorentz
invariant, as it should be. The volume $V$ in (\ref{33}) can be
understood as a mean value of the fireball volume at the stage,
where on the one hand, the hadrons are already formed, i.e. the mean
distances  between them are larger than the confinement  radius
$R_c\sim 1/m_{\rho}\sim (1/4)\,{\rm fm}$, but on the other hand, hadron
interactions are  still  essential and the mean distances  between
the antinucleons are of order or larger than  the deuteron size, so
$\bar{d}$ could be formed in $\bar{p}\bar{n}$ collisions.
 The antinucleon  distributions $n_{\bar{p}}({\bf r}),n_{\bar{n}}({\bf r})$ 
inside the  fireball are essentially nonuniform: at the
stage of the fireball evolution, preceding the dense gas stage, 
antinucleons strongly annihilated in the internal part of the fireball and in much less extent in its
external layer of the thickness of order $\bar{p}(\bar{n})$
annihilation length $l_{ann}$. For this reason
$\ov{n^2}_p/\bar{n}^2_p$ may be essentially larger than 1.
For the same reason the antinucleons and antideuterons from the
backside of the fireball (with respect to the  observer) are absorbed
in the fireball and cannot reach detector, see Fig.~\ref{fig_fire_b}. Therefore,
only one half of the fireball volume contributes to the number of
registered $\bar{p}$, $\bar{n}$ and $\bar{d}$. The corresponding
factor approximately equal to 2 is accounted in (\ref{32}).

The width $\Gamma$ can be calculated in one or another model of fireball
evolution. However, since it enters in the expression for the coalescence 
parameter $B^{th}_2$ (\ref{33}) as $\sqrt{\Gamma}$, it
influences $B^{th}_2$ not significantly. The same remark refers to not
quite certain patameter $a$. Therefore, the comparison with the
data allows one to find the most essential parameter
$V(\bar{n}_{\rho})^2/\ov{n^2_p}$, which would make it possible
to check various models of the fireball evolution.

\section{The model of dense gas stage of fireball evolution.}

We accept the following model for the dense gas stage of
fireball evolution \cite{16}. (A related model had been suggested
long ago \cite{18,19}: it may be called Fermi--Pomeranhuk model).
Neglect for a moment contributions of all particles except for
nucleons and pions. Assume that at dense gas stage any participant
-- nucleon or pion occupies the volume $v_N$ or $v_{\pi}$,
respectively. Then
\be
 n_N\, =\, {N_N\over V}\, =\, {n^0_N\over 1 + Q_{\pi} \beta} \; , \qquad
 n_{\pi}\, =\, {N_{\pi}\over V}\, =\,{n^0_N Q_{\pi}\over 1 + Q_{\pi}\beta}
\label{34} \ee
where $n^0_N = 1/v_N$, $\beta = v_{\pi}/v_N$,
$Q_{\pi}=N_{\pi}/N_N$. For numerical estimations we take
$\beta=(r_{\pi}/r_N)^3=0.55$, where $r_{\pi}=0.66\,{\rm  fm}$ and
$r_N=0.81\,{\rm fm}$ are pion and nucleon electric radii. We take $n^0_N$
as a parameter varying in the interval from $n^0_N=0.17\,{\rm fm}^{-3}$
(normal nuclear density) to $0.30\,{\rm fm}^{-3}$.

Check first the applicability of our approach. Choose $n^0_N=0.24\, {\rm fm}^{-3}$
 and $Q_{\pi}=5.2$. (The latter value was found at NA44
experiment \cite{Af} at SPS.) We have $n_N=0.062\,{\rm fm}^{-3}$,
$n_{\pi}=0.325\,{\rm fm}^3$, $n=n_N+n_{\pi}=0.39\,{\rm fm}^{-3}$ and the mean
distances between the fireball constituents is $d=1/n^{1/3}=1.4\,{\rm fm}$.
Evidently, the conditions $\lambda_{\bar{\pi}}=1/p_{\bar{\pi}}\ll d$ and
$R_c\ll d$ are well satisfied. Check now the balance
conditions  that the probability of deuteron desintegration
exceeds the fireball expansion rate. The former is given by
$2\Gamma(m/E_{\bar{p}})$.  In calculation of $\Gamma$ we assume,
that the hadronic spectra at the dense gas stage of fireball
evolution differ not too much from the spectra at the final
(freeze-out) stage. Since, as was explained above, rather crude
estimation of $\Gamma$ is sufficient for our purposes, we believe,
that such approximations is satisfactory.  Using (\ref{25}) and
the pion  spectrum, presented in \cite{21}, $\Gamma$ was found,
$\Gamma\approx 200$ MeV (Glauber correction was accounted). The
estimation for the fireball expansion is $w\sim(1/5)\,{\rm fm}^{-1}$. We
have: $2\Gamma(m/E_{\bar{p}})\approx 1.4\,{\rm fm}^{-1}\gg 0.2\,{\rm fm}^{-1}$.
$(E_{\bar{p}}\approx 1.5\, {\rm GeV}$ at NA44 experiment
\cite{1}). So, this condition is also fulfilled. Finally, check the
condition ii) of Section 3, which is equivalent to the requirement
$Im\, f\ll d$. Before taking into account the Glauber correction
$Im\, f\approx 1\,{\rm fm}$ and this condition is not well satisfied. After
accounting  30\% Glauber correction the situation improves, but
not too much. For this reason the values of $\Gamma$, presented
above, have a large (may be 50\%) uncertainty. Since
$\sqrt{\Gamma}$ enters (\ref{33}), the error in $B^{th}_2$ reduces
to 25\%.

\section{Comparison with data on {\boldmath $\bar{d}$}-production.}

The antideuteron production in heavy ion collisions was observed
in NA44 experiment at CERN in $Pb+Pb$ collisions at $\sqrt{s}=17.4A\, {\rm GeV}$
\cite{1}, STAR experiment at RHIC in $Au+Au$ collisions at
$\sqrt{s}=130A\, {\rm GeV}$ \cite{2} and E864 experiment at AGS in $Au+Pb$
collisions at $\sqrt{s}=4.8A\,{\rm GeV}$ \cite{3}. The data of these
experiments are presented in Table 1.

\begin{table}
\hspace{22mm}
\begin{tabular}{|c|c|c|c|}\hline
Experiment & NA44 \cite{1} & STAR \cite{2} & E684 \cite{3} \\ \hline \hline
$\sqrt{s}/A$, GeV & 17.4 & 130 & 4.8 \\
$\bar{p}_{\bar{p}\perp }$, GeV & 0.55 & 0.33 & 0.17 \\
$\bar{E}_{\bar{p}\, cm}$, GeV & 1.5 & 1.05 & 0.99 \\
\# of "wounded" nucleons & 362 \cite{20,21} &  320 \cite{22} & 350 \cite{23} \\
$Q_{\pi}$ & 5.2 \cite{20,21} & $7\pm 1$ \cite{24} & 1.6 \cite{23}\\ \hline
\end{tabular}
\caption{Experiments, where antinuclei were observed}
\end{table}

Table 2 gives the values of coalescence parameters $B_2^{exp}$,
measured in these experiments and relevant centralities.
It must be mentioned that $B^{exp}_2$ rather strongly depends on the centrality.
For example in NA44 experiment the results for 0--5\% centrality are about 1.5 times lower.  Also,
$B^{exp}_2$ has significant $p_{pt}$ dependence:
such dependence  was observed at PHENIX experiment at RHIC
(we do not analyse PHENIX data here, since only preliminary results were published
up to date \cite{4}).

Table 2 presents also the results of  our calculation for the
values of $n^0_N$, introduced in (\ref{34}): $n^0_N=0.17, 0.24, 030\, {\rm fm}^{-3}$.
In the calculation of mean volume $V$ at the dense gas stage the number
of  "wounded" nucleons and the number of pions were taken from
Table 1, the corrections 15\%, 20\% and 10\% were accounted for other particles,
except for nucleons and pions, in cases of NA44, STAR and E684 experiments respectively. The
antiproton annihilation length was estimated as
$\bar{l}_{\bar{p},ann}\approx 3\, {\rm fm}$ in case of NA44 and STAR
experiments and  $l_{\bar{p},ann}\approx 1.5\,{\rm fm}$ in case of E684. In
all cases we can put with sufficient accuracy
$\ov{n^2}/\bar{n}^2\approx 2$ and $a=1/2$.

\begin{table}
\hspace{10mm}
\begin{tabular}{|c|c|c|c|c|}\hline
\multicolumn{2}{|c|}{Experiment} & NA44  & STAR & E684 \\ \hline
\multicolumn{2}{|c|}{Centrality, \%} & 0--10 & 0--18 & 0--10 \\
\multicolumn{2}{|c|}{$B^{exp}_2$, $10^{-4}\, {\rm GeV}^2$}&
 $4.4\pm 1.3$ & $4.5\pm 0.3\pm 1.0$ &$41\pm 29\pm 23$\\ \hline\hline
 & $\Gamma$, MeV & 148 & 159 & 109 \\
$n^0_N=0.17\,{\rm fm}^{-3}$ & $V$, $10^3\,{\rm fm}^3$ & 8.51 & 10.95 &4.26 \\
 & $B^{th}_2$, $10^{-4}\,{\rm GeV}^2$ & 3.17 & 3.38 & 11.2 \\ \hline
 &  $\Gamma$, MeV & 187 & 206 & 142 \\
$n^0_N=0.24\,{\rm fm}^{-3}$ & $V$, $10^3\,{\rm fm}^3$ & 6.03 & 7.76 & 3.02 \\
 & $B^{th}_2$, $10^{-4}\,{\rm GeV}^2$ & 4.00 & 4.22 & 13.9 \\ \hline
 & $\Gamma$, MeV & 214 & 232 & 187 \\
 $n^0_N=0.30\,{\rm fm}^{-3}$ & $V$, $10^3\,{\rm fm}^3$& 4.26 & 6.21 & 2.42 \\
 & $B^{th}_2$, $10^{-4}\,{\rm GeV}^2$ & 5.28 & 4.96 & 15.0 \\ \hline
\end{tabular}
\caption{Antinucleon width, fireball volume and antideuteron coalescense parameters}
\end{table}

The calculated values of the coalescence parameters $B^{th}_2$ are
presented in Table 2. It must be mentioned, that in case of E684 experiment the validity conditions
of our approach are on the edge of their applicability. So, the
theoretical expectations for $B_2$ in this case are valid only by
the order of magnitude. In two other cases the agreement of the
theory with the experiment is quite satisfactory. Let us remark,
that the main uncertainty arises from the effective width
$\Gamma$ which, probably, is known with an accuracy of order
50\%. Taking in mind all other possible uncertainties, we believe
that the accuracy of theoretical predictions for $B_2$ is also
about 50\%. The inspection of the Table 2 shows, that the total
hadronic density (pions+nucleons) at the dense gas stage of
fireball expansion, where the antideuterons are formed, is of
order $n_N+n_{\pi}\sim 0.4\, {\rm fm}^{-3}$. Much lower or much higher
densities would lead to contradiction with experiment.

Using the pion and nucleon densities and spectra chosen above,
we can estimate the typical energy density at the dense gas stage:
$\epsilon \sim 0.3 \div 0.5\,{\rm GeV}/{\rm fm}^3$. As the colliding ions are
initially strongly Lorentz-compressed and the produced fireball expands
in the longitudinal direction with an almost velocity of light, it takes
approximately $R\approx 10\,{\rm fm}$ to reach the proposed volume. It is instructive
to put these values on the Fig.~27 from \cite{McL},
where the density of energy/time evolution of the fireball is represented.
We see that the dense gas stage begins just after the moment when
the total hadronization happened and this is
a good check for self-consistency of our assumptions.

\section{Production of {\boldmath $^3\ov{He}$}.}

The calculation of $^3\ov{He}$ production proceeds along the
same lines as $\bar{d}$. The only difference is that now the
collision integral $I$ -- the first term in the r.h.s.~of eq.~(\ref{20})
corresponds to the formation of $^3\ov{He}$ in collision of three
antinucleons: $\bar{p_1},\bar{p_2}$ and $\bar{n}$ and the effective coupling constant is given by
(\ref{19}). Instead of Breit-Wigner off-shell mass distributions for antiparticles
$\bar{p_1},\bar{p_2},\bar{n}$ and $^3\ov{He}$   we now take the Gaussian ones:
\be
f_i(m'_i)\,=\,{2\over \sqrt{\pi}\,\Gamma_i}\, e^{-(m'_i-m_i)^2/(\Gamma/2)^2}
\label{35}
\ee
The reason is that the Breit-Wigner distributions do not provide the necessary
convergence of collision integral.
The collision integral takes the form:
\bea
{I\over E_{He}} & = & {\pi \,G^2 \over 64\, E_{He}} \int \Bigl\{ \prod_{i=1,2,3}  dm'_i \, f_i(m'_i)\Bigr\}
\, \int {d^3 p_1\over E'_{p_1}}\,
{d^3 p_2\over E'_{p_2}} \, {d^3 p_n\over E'_{p_n}}\,q_p(p_1)\, q_p(p_2)\, q_n(p_n) \nonumber \\
 & & \times \, \delta^3(p_1+p_2+p_n-p_{He})\, \delta(E'_{p_1} +E'_{p_2}+E'_{p_n} -E'_{He})
\label{36}
\eea
(The distribution of $^3\ov{He}$ was taken as a $\delta$-function, what is
equivalent to $a=\tilde{\Gamma}_{He}/(3\Gamma) \ll 1$). The integral is equal to:
\be
I\,=\, {\pi^4\over 96\sqrt{6}} \,G^2\, \Gamma^2 \, q^3(p)
\label{37}
\ee
Applying the balance condition and (\ref{19}) we find the coalescence  parameter:
\be
B_3\, =\,96\pi^7\, {1\over \sqrt{2}} \, {\Gamma\over \Lambda}\, {1\over V^2}
\,{\ov{n^3}\over \bar{n}^3}
\label{38}
\ee
The accuracy of calculation of $B_3$ is lower, that in case of $B_2$
since additional uncertainties appear: strong dependence on
 ultraviolet cut-off $\Lambda$ etc. So we may pretend only on the estimation of $B_3$,
correct up to order of magnitude (in the best case up to factor of 2).
The coalescence parameter for $^3\ov{He}$ production in $Au+Au$ collisions at
$\sqrt{s}=130A\, {\rm GeV}$ was measured by STAR Collaboration \cite{2}.
In this experiment the centrality  was up to 18\%,
$^3\ov{He}$ transverse momentum $1.0 < p_{\perp, \ov{He}}< 5.0\,{\rm GeV}$
and rapidity $| y | < 0.8$. These   limititations corresponds to the average
antiproton energy $E_{\bar{p}}\approx 1.4\,{\rm GeV}$. STAR found:
\be
B^{exp}_{3, \, \ov{He}}\,=\,(2.1 \pm 0.6 \pm 0.6)\times 10^{-7}\,{\rm GeV}^4
\label{39}
\ee
The theoretical value according to (\ref{38}) at $n^0_N=0.24\,{\rm fm}^{-3}$,
$\ov{n^3}/\bar{n}^3=3$, and $\Lambda=300\,{\rm MeV}$ is:
\be
B_{3,\,\ov{He}} \,=\,3.3\times 10^{-7}\,{\rm GeV}^4
\label{40}
\ee
The agreement with experiment is good, despite of many theoretical
uncertainties. It demonstrates the validity of basic
ideas of theoretical approach.

\section{Summary and conclusion.}

The coalescence parameters for production of antideutrons and
antihelium-3 in heavy ion collisions were calculated. The obtained
results are based on three assumptions: i) the  main mechanism of
light antinucleous  production is coalescence (fusion) mechanism
--- eq.~(\ref{1},\ref{2}); ii) all particles, participating in
fusion  process are moving in the mean field of other fireball
constituents; iii) the number of produced antinucleous is
determined by the balance conditions: the equality of produced  and
desintegrated --- mainly by pions --- antinuclei.  The production of
antinucleous proceeds at the dense gas stage of the fireball evolution,
when the hadrons are already formed, but their interaction is still
important. Statistical or thermal equilibrium are not supposed at the dense gas stage.
In fact, the final results depend on one parameter ---
the volume of the fireball at this stage (or, equivalently, on the hadron densities.)
Good  agreement with experimental data
for coalescence parameters was obtained for experiments at CERN,
RHIC and AGS for the values of $n^0_N$, defined by eq.~(\ref{34}),
$n^0_N\sim 0.17 \div 0.30\,{\rm fm}^{-3}$, close to normal nucleous
density. Much lower, or much higher values of $n^0_N$ lead the to
values of  coalescence parameters, incompatibe with
the data. It must be stressed, that the same values of $n^0_N$
well describe the coalescence parameter for $^3\ov{He}$ production,
which demonstrates the effectiveness of the method. The
values of the fireball volume $V$ are about 2 times larger, than those
found in \cite{13} at the so-called chemical freeze-out stage, about
2 times smaller than at thermal freeze-out \cite{24,25} and in
agreement with limitations on the volumes found in \cite{14}. The
same method can be applied to the production of deuterons and $^3He$
nucleous. But in these cases, in order to avoid the background,
consisting of $d$ or $^3He$ falling apart from colliding nuclei,
it is necessary to select the events with rapidity close
to zero. One may expect, that for such events the expressions for
coalescence parameters, found above are valid. The further
investigations of this problem --- both theoretical and
experimental are very desirable, since they can shed light also on
the most interesting stage --- the stage of hadron formation.

\section*{Acknowledgements}

We are thankful to G.~Brown. L.~McLerran, E.~Shuryak for
discussions and S.~Kiselev, Yu.~Kiselev, A.~Smirnitsky, N.~Rabin for
information about experimental data.

This work was supported in part by INTAS grant 2000-587 and RFBR
grant 03-02-16209.

\section*{Appendix A: Account of the finite nuclear force radius}

\setcounter{equation}{0}
\def\theequation{A.\arabic{equation}}

According to Bethe the general form of $pn$ scattering phase
$\delta$ at low energies is
\be
k\,{\rm ctg}\, \delta_0 =-\alpha+\frac{1}{2}(\alpha^2+k^2)r_0
\ee
Here $k$ is the proton (or neutron) momentum in c.m.~frame, $r_0$ is defined as
a radius of nuclear forces: at $r >r_0$ the $pn$ potential
$V_{pn}(r)$ may be neglected. (A.1) is an expansion in terms of
$kr_0^2$, higher order terms, $\sim k^4 r_0^3$ are omitted. For
the scattering amplitude (\ref{7}) we get from (A.1)
\be
f_0\,=\,{1\over k({\rm ctg}\,\delta_0 -i)}\,=\,
{i\over k-i\alpha}\,{1\over 1+i(k+i\alpha)r_0/2}
\ee
The value of effective coupling constant is determined by the residue of the
amplitude $f$ at the pole $k\to i\alpha$. It is easy to see, that
in comparison with the case $r_0=0$, the value of the residue
changes by the factor $(1-\alpha r_0)^{-1}$. Higher order
corrections $\sim k^4 r^2_0$ are negligible.

There are several different definitions of the nuclear force radius
$r_0$. On the other hand, all $r_0$-corrections to the $pnd$-coupling can be
expressed in terms of the normalization constant $A_S$ of the deuteron radial wave function
$R=A_S e^{-\sqrt{m\ve}r}$ at large $r$. Indeed, it determines the residue of the amplitude
(\ref{7}) at the pole $E=-\ve$:
\be
\label{f0amp}
f_0\,=\,-\,{A_S^2\over m} \,{1\over E+\ve}  \; \qquad (E\to -\ve)
\ee
In case of zero radius of nuclear force $A_S=(4m\ve)^{1/4}=0.68\,{\rm fm}^{1/2}$.
Analisys of experimental data on elastic $pd$ scattering
\cite{BGK} as well as various potential models give the value around $A_S=0.88\,{\rm fm}^{1/2}$.
Comparison with the amplitude (\ref{7}) gives the following coupling value:
\be
g^2\,=\,64\pi m A_S^2
\ee
It exceeds the zero radius result (\ref{9}) by $1.67$ times. It corresponds to $r_0=1.7\,{\rm fm}$.

\section*{Appendix B: Estimation of direct process {\boldmath $NN\to d\pi$}}

\setcounter{equation}{0}
\def\theequation{B.\arabic{equation}}

\begin{figure}[tb]
\hspace{45mm}\epsfig{file=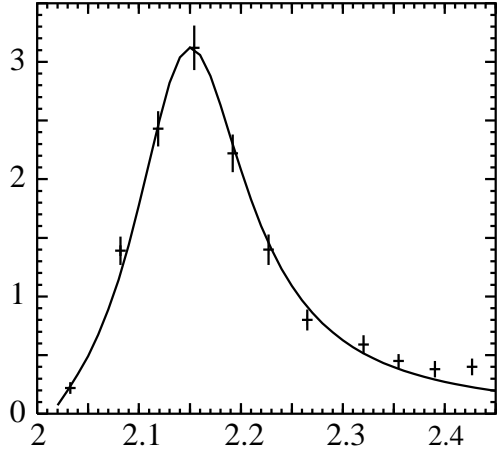, width=65mm}
\caption{Total cross section $\sigma_{pp\to d\pi^+}$ in mb versus $\sqrt{s}$ in GeV. The experimental data
are taken from \cite{SHM}. The line represents the fit of  \cite{VA}.}
\label{fig_ppdpi}
\end{figure}

The main processes of the deutron formation in vacuum is $NN\to d\pi$.
In case of exact isospin symmetry the total cross sections of each chanell are related as
\be
 \sigma_{pp\to d\pi^+}\,=\,\sigma_{nn\to d\pi^-}\,=\,2\,\sigma_{pn\to d\pi^0} \; ,
\label{diso}
\ee
similarly for antinucleons. The process $pp\to d \pi^+$, as well as inverse one,
has been accurately measured in many experiments.
The cross section is shown in Fig \ref{fig_ppdpi}. It has a resonanse peak near $\sqrt{s}=2.17\, {\rm GeV}$,
the mass of the $N\Delta$ system and decreaces  steeply above $s \sim 2.5 \, {\rm GeV}$.

In this appendix we estimate the contribution of these processes to the deutron formation rate.
It is convenient to use invariant definition of the cross-section in relativistic calculations. We assume
$q_p(p)=q_n(p)$. The processes with identical particles in initial state
$pp\to d\pi^+$, $nn\to d\pi^-$ should be taken with weight $1/2$ to avoid the double counting
from the same regions of the phase space. So the total deuteron production rate can be written as:
\be
\label{trpdpi}
{m_d\over E_d} \, u^\mu_d {\partial q_d(p_d)\over \partial x^\mu} \,=\,{3\over 2}\,
\int  \, d^3p_1 \, d^3p_2 \, d^3p_\pi \, q_p(p_1) \, q_p(p_2) \, \omega_{pp\to d\pi^+}\,=\,{I_{direct}\over E_d}\; ,
\ee
where $\omega$ is the reaction probability:
\be
\omega_{pp\to d\pi^+}\,=\,{\sqrt{(p_1p_2)^2-m_p^4}\over E_1 E_2} \,
{d\sigma_{pp\to d\pi^+}\over d^3 p_d \,d^3p_\pi}
\ee
The value $I_{direct}$ should be compared with the collision integral $I$ of the fusion process,
obtained in (\ref{ibw},\ref{24}).  Integrating by the momenta $p_\pi$, one obtains:
\be
I_{direct}\,=\,{3\over 2}\, \int  \, {d^3p_1\over E_1} \, {d^3p_2\over E_2} \, q_p(p_1) \, q_p(p_2) \,
\sqrt{(p_1 p_2)^2-m_p^4} \, E_d {d^3 \sigma_{pp\to d\pi^+}\over d^3 p_d}
\label{idir}
\ee
We compute this integral in the center of mass frame. Let $({\cal E}, {\bf k})$ and $({\cal E}_d, {\bf k}_d)$
be the c.m.~energies and momenta of the first proton and the deuteron respectively.
Then the lab momenta are given by the Lorentz boost, ${\bf v}$ is the center of mass velocity:
$$
{\bf p}_1\,=\;\;\; {\bf k} + (\gamma-1) {{\bf v}({\bf v}{\bf k})\over v^2} + \gamma {\bf v} {\cal E}
 \; , \qquad  E_1\,=\,\gamma ( {\cal E}+{\bf v}{\bf k})
$$
\be
{\bf p}_2\,=\, -{\bf k} - (\gamma-1) {{\bf v}({\bf v}{\bf k})\over v^2} + \gamma {\bf v} {\cal E} 
 \; , \qquad  E_2\,=\,\gamma ( {\cal E}- {\bf v}{\bf k})
\ee
where $\gamma=(1-v^2)^{-1/2}$. Now in the integral (\ref{idir}) we come from the variables
${\bf p}_1,{\bf p}_2$ to ${\bf v}, {\bf k}$ according to:
$$
{d^3 p_1\over E_1}\,{d^3 p_2\over E_2}\,=\,8\gamma^4 {\cal E} \,d^3 v\, d^3k
$$
Notice, that $Ed^3\sigma/d^3p$ is invariant cross-section, so 
$E_d d^3\sigma/d^3p_d={\cal E}_d d^3\sigma/d^3k_d$ . Then the integral (\ref{idir}) can be written as:
\be
\label{idir2}
I_{direct} \,=\, {3\over 2} \cdot 16 \int   \,d^3v \, d^3k\, q_p(p_1)\,  q_p(p_2) \, 
  \, \gamma^4 {\cal E}^2 k\, {\cal E}_d {d^3\sigma_{pp\to d\pi^+}\over d^3 k_d} \, 
\ee
And finally, it could be convenient instead of the velocity ${\bf v}$ to substitute the 
deuteron cm momentum ${\bf k}_d$ according to:
\be
{\bf v}\,=\,{2{\bf w}\over 1+{\rm w}^2}    \; , \qquad {\bf w}\,=\,{{\bf p}_d-{\bf k}_d\over E_d+{\cal E}_d}
\ee
The velocity integration measure becomes:
\be
\gamma^4\, d^3v\,=\,{8\over (1-{\rm w}^2)^3}\,d^3{\rm w}\,=\,\left( {E_d+{\cal E}_d\over 
m_d^2+E_d{\cal E}_d+{\bf p}_d{\bf k}_d } \right)^2 {d^3 k_d\over {\cal E}_d}
\ee
Then the integral (\ref{idir2}) can be written as follows: 
\be
\label{idir3}
I_{direct} \, =\, {3\over 2}\cdot 16 \int  d^3k \; d^3k_d \; {d^3\sigma_{pp\to d\pi^+}\over d^3 k_d} \; 
{ k{\cal E}^2 \, ( E_d+{\cal E}_d)^2 \over ( m_d^2+E_d{\cal E}_d+{\bf p}_d{\bf k}_d )^2 } \; q_p(p_1) \, q_p(p_2)
\ee
To evaluate it further, one needs to know the antiproton distribution $q_p(p)$ in the fireball.
For the first approximation, we could take $q_p(p)\approx q_p(p_d/2)$, aw we did in (\ref{24}).
This however would be an overestimation of the production rate, since 
it gets the main contribution from the resonace area (see Fig.~\ref{fig_ppdpi}), but not from the
threshold. More carefull estimation can be obtained by taking 
the Boltzman distribution for the antiprotons:
\be
\label{boltd}
q_p(p)\,\sim \, e^{-E/T} 
\ee
Then one can perform the integration over the angles and $k_d$. The result is: 
\bea
I_{direct}& = &  q_p^2(p_d/2)\, {3 \pi\over 4 m_d^2}\,\int_{(m_d+m_\pi)^2}^\infty 
 ds \,\cdot\, s (s-4m_p^2)  \, \sigma_{pp\to d\pi^+} \nonumber \\
 & & \times \exp{ \left(-\, {E_d(s-m_d^2-m_\pi^2)\over 2Tm_d^2} \right)} \,
 \, {{\rm sh}\!\left( \sqrt{s}p_d k_d/(Tm_d^2) \right)\over \sqrt{s}p_d k_d/(Tm_d^2) }
\label{idir4}
\eea
where $\sigma_{pp\to d\pi^+}$ is the total cross section,
$k_d^2=[s-(m_d-m_\pi)^2][s-(m_d+m_\pi)^2]/(4s)$.

The integral (\ref{idir4}) reaches maximum for the deuteron at rest in the
frame, where the system is described by Boltzman distribution (\ref{boltd}). So for numerical
estimation we take $p_d=0$ and the temperature $T$ of order of the antiproton inverse slope
parameter $T=300\,{\rm MeV}$. We intergate (\ref{idir4}) up to $\sqrt{s}=2.5\,{\rm GeV}$.
Above this value other channels appear, where experimental data are rare.
However the total cross section $pp\to dX$ at high $\sqrt{s}$ is negligible,
so this will not be an underestimation. The result is
\be
I_{direct}\,=\,7.5\,{\rm GeV}^2 \times q_{\bar p}^2(p_{\bar d}/2)
\ee
It should be compared with the fusion rate integral (\ref{24}). According to our estimations
$I=40\,{\rm GeV}^2 \times q_{\bar p}^2(p_{\bar d}/2)$. It confirms our assumption that the contribution
of direct process of antideuteron formation ${\bar N}{\bar N}\to {\bar d}\pi$ is small
in the heavy ion collisions.

\section*{Appendix C: Particle width in dense medium}

\setcounter{equation}{0}
\def\theequation{C.\arabic{equation}}

The relation (\ref{22}) for $\Gamma_{\bar p}$
corresponds to the case of nucleon moving through the medium of
pions at the dense gas stage of fireball evolution. It turns out that
$\Gamma_{\bar p}$
weakly depends on the pion spectrum details, so with sufficient accuracy one can use
the observed pion spectrum which is formed on the freeze-out stage.
To obtain numerical results we exploit eq.~(\ref{gammad}) for pion-antiproton
collisions, where the pion distribution is extracted from \cite{21}.
It is usually parametrized in terms of transverse mass $m_\pi^{\perp}=\sqrt {(p_\pi^{\perp})^2 +m^2_\pi}$
and rapidity $y={1\over 2} \ln [(E+p^\parallel)/(E-p^\parallel)]$:
\be
2\pi E_\pi q_\pi (p) =
\frac{d^2 n_{\pi}}{m_\pi^{\perp}dm_\pi^{\perp} dy} =
C\exp\{-(m^{\perp}_\pi-m_\pi)/T_\perp-y^2/b^2\},
\ee
where $T_\perp=110\,{\rm MeV}$, $b=1.9$ and $C\approx 38.6\,{\rm GeV}^{-2}$ is the normalization constant,
determined from the requirement $\int q_\pi (p) d^3p=n_\pi$.
Performing the calculation we find that the numerical value of $\Gamma_{\bar p}$
is about $290\,{\rm MeV}$ at the pion density $n_\pi=0.32\,{\rm fm}^{-3}$ and antiproton energy
$E_{\bar p}=1.5\,{\rm GeV}$. The effect of averaging over pion momentum is
practically negligible as the obtained value of $\Gamma_{\bar p}$
differs by less than $5\%$ from
its value calculated within the assumption that all pions are at rest.

As it was pointed in the text, Eq.~(\ref{22}) implies that the scattering amplitude
of the incident particles on the medium constituents is much less
then the mean distance
between them and the scattering can be treated as the sum of independent
processes. However, this condition is not fulfilled quite well in the dense matter.
In this case $n\sigma$ has the meaning of the total cross section on
all the constituents of fireball (divided by its volume) which is not simply
reduced to the sum of individual cross sections. This lack of additivity is
well known from the nucleon-deuteron scattering at high energies \cite{Gl}.
Within the accuracy of our approach it is sufficient to take only the first term
in density expansion. Consider the scattering amplitude at small angles
when the incident nucleon interacts with the target consisting from two pions
confined in a cube with side $a$.
From the calculation of $\Gamma_{\bar p}$ it
follows that the main contribution to the integral for $\Gamma_{\bar p}$ arises
from the pions with small momenta, so one can assume that the antiproton momentum
is much larger than the pion momentum. Then the antiproton scattering amplitude
can be written in the framework of method developed by Glauber \cite{Gl}:
\be
f({\bar q})=\frac{k}{2\pi i} \int d^2 \bar \rho e^{-i {\bar q}{\bar \rho}}
\frac{d^3{\bar r_1}d^3{\bar r_2}}{a^6}
[\exp\{2i\delta_1 (\bar \rho - \bar r^{\perp}_1)
+2i\delta_2 (\bar \rho -\bar r^{\perp}_2)\} -1],
\ee
where $\delta_i$ is the phase shift induced by the scattering of antiproton
on $i$-th pion and $r_i^{\perp}$ is the transverse coordinate of $i$-th pion.

Using the relation
\be
\exp\{2i\delta(r)-1 \}=\frac{2\pi i}{k} \int f_i(\bar q)
e^{i {\bar q}{\bar \rho}} \frac{d^2 \bar q}{{(2\pi)^2}}
\ee and identity $e^{(x+y)}-1=(e^x-1)+(e^y-1) + (e^x -1)(e^y -1)$,
we get the following equation for the forward scattering amplitude:
\be
f(0)=f_1 (0) +f_2 (0) + \frac{2\pi i}{k a^2} f_1 (0) f_2 (0).
\ee
To obtain quantitative result we assume that $f_1=f_2$  and $Re\,f_i=0$
(actually, $Re\, f/Im\, f \approx20\%$) and note that $n=2/a^3$
which gives the first correction to $n\sigma$:
\be
n_\pi \sigma \rightarrow n_\pi \sigma \left[1-\frac{\sigma_{\pi \bar p}}{4a^2}
\left(\frac{n_\pi}{2}\right)^{2/3}\right]
\label{screening}
\ee
The factor in square brackets represents the screening effect and also appears
in $\Gamma_{\bar p}$. At the pion density $n_\pi=0.32\,{\rm fm}^{-3}$ and
$\sigma_{\pi \bar p}\approx 30\,{\rm fm}^2$
the screening effect reduces the value of
$\Gamma_{\bar p}$ to $190\,{\rm MeV}$. (It is worth to note that the screening correction
($\ref{screening}$)
to $\Gamma_{\bar d}$ is larger than the same one to $\Gamma_{\bar p}$ as
$\sigma_{\pi \bar d}>\sigma_{\pi \bar p}$
which is a direct indication that $\Gamma_{\bar d}<2\Gamma_{\bar p}$.)
Taking different values of the pion density
one can find the other values of $\Gamma_{\bar p}$ presented in Table 2.

\end{document}